\documentclass{article}

\usepackage{microtype}
\usepackage{graphicx}
\usepackage{subcaption}
\usepackage{booktabs} 
\usepackage{formaljitterbug}
\usepackage{fancyvrb}
\usepackage{color}
\usepackage{amsmath}
\usepackage{amssymb}
\usepackage{mathtools}
\usepackage{amsthm}
\usepackage{listings}
\usepackage{fancyvrb}
\usepackage{tabularx}
\usepackage{makecell}
\usepackage{todonotes}
\usepackage{minted}
\usepackage[T1]{fontenc}

\usepackage{xurl}

\definecolor{keywordcolor}{rgb}{0.7, 0.1, 0.1}   
\definecolor{tacticcolor}{rgb}{0.0, 0.1, 0.6}    
\definecolor{commentcolor}{rgb}{0.4, 0.4, 0.4}   
\definecolor{symbolcolor}{rgb}{0.0, 0.1, 0.6}    
\definecolor{sortcolor}{rgb}{0.1, 0.5, 0.1}      
\definecolor{attributecolor}{rgb}{0.7, 0.1, 0.1} 

\usepackage{hyperref}
\usepackage[capitalize,noabbrev]{cleveref}

\usepackage[preprint]{icml2026}

\begin{document}

\icmltitlerunning{ATLAS: Automated Toolkit for Large-Scale Verified Code Synthesis}

\twocolumn[
\icmltitle{ATLAS: Automated Toolkit for Large-Scale Verified Code Synthesis}

\icmlsetsymbol{equal}{*}

\begin{icmlauthorlist}
\icmlauthor{Mantas Bakšys}{univ,comp}
\icmlauthor{Stefan Zetzsche}{comp}
\icmlauthor{Olivier Bouissou}{comp_us_boston}
\icmlauthor{Remi Delmas}{comp_us_boston}
\icmlauthor{Soonho Kong}{comp_us}
\icmlauthor{Sean B. Holden}{univ}
\end{icmlauthorlist}

\icmlaffiliation{univ}{University of Cambridge, Cambridge, UK}
\icmlaffiliation{comp}{Amazon Web Services, London, UK}
\icmlaffiliation{comp_us_boston}{Amazon Web Services, Boston, USA}
\icmlaffiliation{comp_us}{Amazon Web Services
Santa Clara, CA, US}

\icmlcorrespondingauthor{Mantas Bakšys}{mb2412@cam.ac.uk}
\icmlcorrespondingauthor{Stefan Zetzsche}{stefanze@amazon.co.uk}

\icmlkeywords{Machine Learning, Theorem Proving, Dafny, LLMs, Formal Methods}

\vskip 0.3in
]



\begin{abstract}
Large language models have become proficient at generating functional code, but ensuring the output truly matches the programmer's intent remains difficult. Testing improves trust, yet for safety-critical applications, formal verification provides the only true guarantees through machine-checked proofs. However, verified code remains scarce compared to mainstream languages or mathematical theorem proving, limiting LLM capabilities in this domain. We present ATLAS, an automated pipeline that synthesizes verified programs to address this data bottleneck. Applied to the TACO dataset of Python solutions to LeetCode-style problems, ATLAS generates 2.7K verified Dafny programs, each with high-quality specifications and machine-checked proofs. Through task decomposition, we extract 19K training examples. Fine-tuning Qwen 2.5 7B Coder on this data improves performance from 32.4\% to 56.9\% on DafnyBench and from 15.8\% to 65.8\% on DafnySynthesis, demonstrating that synthetic data generation is a viable path to scaling LLM capabilities for formal verification.
\end{abstract}







\printAffiliationsAndNotice{ \ Mantas Bakšys worked on this project during an internship at Amazon Web Services in London, UK. \newline
\underline{ATLAS} - \underline{A}utomated \underline{T}oolkit for \underline{LA}rge-Scale Verified Code \underline{S}ynthesis}

\section{Introduction}

\begin{figure}[t]
    \centering
    \includegraphics[width=\columnwidth]{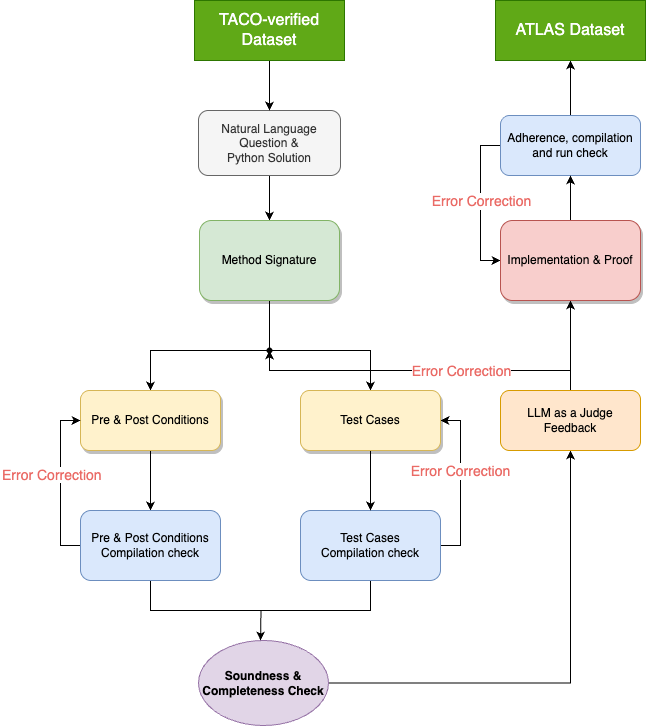}
    \caption{An overview of the ATLAS pipeline.}
    \label{fig:pipeline}
\end{figure}

Large language models (LLMs) have transformed software engineering, demonstrating impressive capabilities in code generation, completion, and debugging. On mainstream languages like Python, JavaScript, and Java \citep{chen2021codex, cassano2023multiple, kocetkov2022stack}, models routinely produce functional, semantically meaningful code that solves the task at hand. Yet in typical workflows, even high-quality generated code comes without guarantees—there is no systematic way to ensure the output truly matches the user's intent. Unit tests and property-based testing can improve trust, and for most applications this is sufficient.

In safety-critical domains such as cryptographic libraries, medical devices, or aerospace systems, however, "probably correct" is not enough. Formal verification addresses this gap: by requiring precise specifications of intended behavior and mathematically proving implementations satisfy them, verified software eliminates entire classes of bugs by construction. Producing verified code today, however, demands substantial expertise—developers must craft not only the implementation but also specifications and the annotations that guide automated reasoning toward a proof.

Formal verification tools span a spectrum. Interactive theorem provers (ITPs) like Lean and Isabelle require users to construct proofs step-by-step and have seen the most success in mathematical domains. AI-assisted theorem proving in ITPs has advanced rapidly—multiple systems recently achieved gold medal performance on IMO 2025 with formal proofs \citep{seed-prover, achim2025aristotle}. For software verification, auto-active tools like Dafny \citep{dafny}, F$^{*}$ \citep{fstar}, Viper \citep{viper}, and Verus \citep{lattuada2023verus} offer a different tradeoff: users annotate code with specifications and proof hints, and an SMT solver automatically discharges the verification conditions. This automation handles routine reasoning, letting human effort focus on the hard parts. We focus on Dafny and provide further background in \Cref{sec:background}.

Can similar AI-driven progress be achieved for auto-active verification as has been seen in ITPs? A key obstacle is data scarcity. While code in mainstream languages is abundant, verified code remains rare—Dafny repositories on GitHub are orders of magnitude fewer, and many contain incomplete proofs. The Lean community addressed analogous challenges through synthetic data: autoformalized theorem statements have successfully trained models for theorem proving \citep{polu2022formalmathematicsstatementcurriculum, ying2025leanworkbooklargescalelean}. Inspired by these successes, we tackle the data scarcity problem for Dafny—though our challenge differs: rather than autoformalizing mathematical statements, we must synthesize entire verified programs comprising implementations, specifications, and proof annotations.

To this end, we introduce ATLAS, a pipeline for automatically generating verified programs at scale (\Cref{fig:pipeline}). Given a natural language problem description, a reference implementation in a mainstream language (e.g., Python), and test cases, ATLAS synthesizes a corresponding program in a verification-aware language, complete with formal specifications and proof annotations. The pipeline works through iterative refinement, alternating between verifier feedback and LLM-driven repair. To ensure specification quality, we verify that specifications are consistent with test cases and, ideally, strong enough to uniquely determine the expected output for each test input. While the pipeline could be adapted to other input datasets and target languages (e.g., Verus, F$^*$) with modest effort, we instantiate it on the TACO-verified dataset \citep{li2023tacotopicsalgorithmiccode}—LeetCode-style problems with Python solutions and unit tests—targeting Dafny.

\begin{figure}[t]
\centering
    \includegraphics[width=\columnwidth]{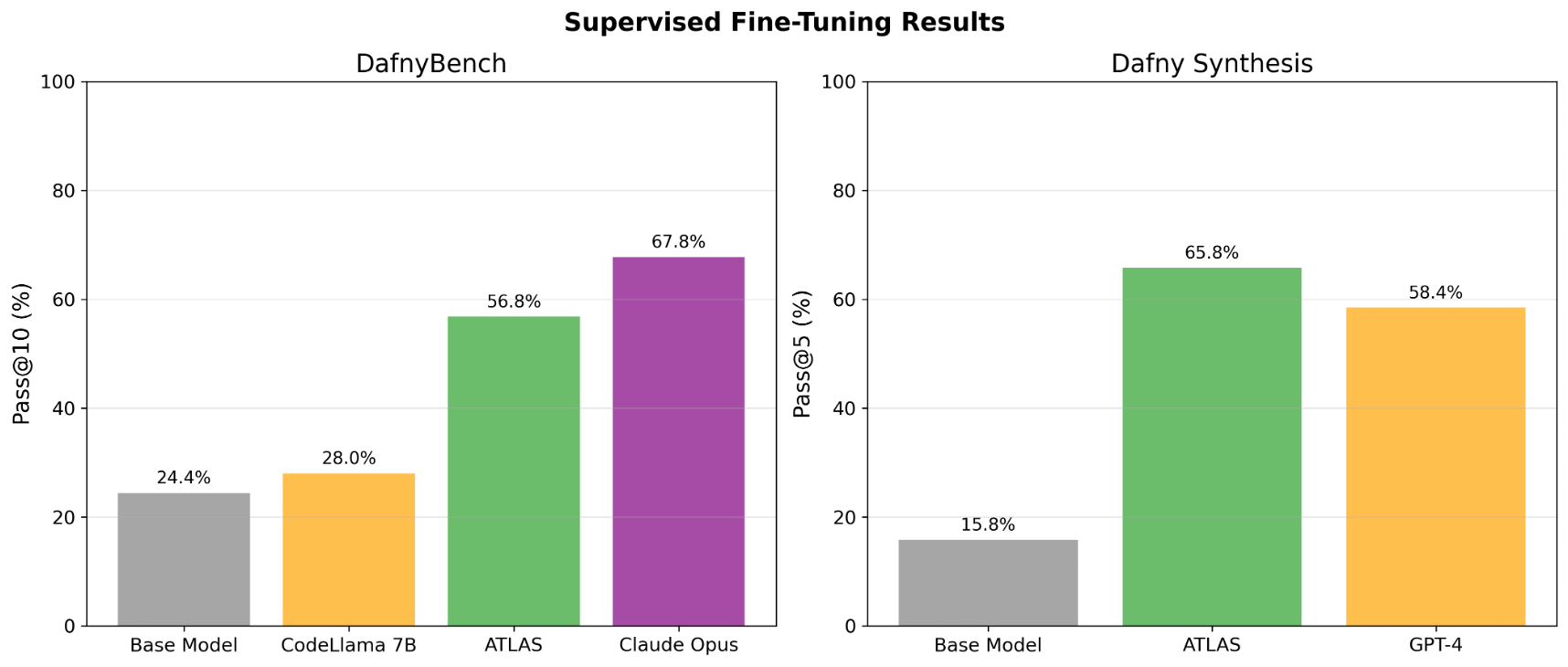}
\caption{Evaluation of our fine-tuned Qwen 2.5 7B Coder on DafnyBench and DafnySynthesis, compared with results reported in \citep{loughridge2024dafnybenchbenchmarkformalsoftware} and \citep{misu2024towards} at the time of their release.}
\label{fig:training-results}
\end{figure}

This yields 2.7K verified Dafny programs, each with high-quality specifications and machine-checked proofs of correctness. By decomposing each into specialized training tasks—such as synthesizing code and specifications from natural language, inferring implementations from specifications, and repairing specifications, implementations, and proofs—we achieve 7$\times$ data expansion for $\approx$19K training examples (\Cref{tab:sft_all_tasks}). Fine-tuning Qwen 2.5 7B Coder on this data improves performance from 32.4\% to 56.9\% on DafnyBench \citep{loughridge2024dafnybenchbenchmarkformalsoftware}, which evaluates proof annotation of existing Dafny code, and from 15.8\% to 65.8\% on DafnySynthesis \citep{misu2024towards}, which evaluates generating verified programs from scratch (\Cref{fig:training-results}).

To sum up, our contributions are the following:
\begin{itemize}
\item We present ATLAS, an automated pipeline for synthesizing verified programs from algorithmic problems with reference implementations and test cases.
\item We define soundness and completeness criteria that leverage test cases to ensure specification quality, filtering degenerate solutions.
\item We demonstrate a data decomposition strategy that extracts multiple training tasks per verified program, achieving 7$\times$ data expansion.
\item We show empirically that fine-tuning on our synthesized data yields substantial improvements on DafnyBench and DafnySynthesis.
\end{itemize}

\section{The Dafny Language}
\label{sec:background}

Dafny is a verification-aware programming language based on Floyd-Hoare logic \citep{hoare1969axiomatic}. Programs are annotated with specifications in the form of Hoare triples $\{P\}\ s\ \{Q\}$, where $P$ is a precondition (\texttt{requires}), $Q$ is a postcondition (\texttt{ensures}), and $s$ is the program body. Figure~\ref{fig:dafny-example} shows a simple example.

\begin{figure}[t]
\begin{lstlisting}[language=dafny, frame=single, basicstyle=\small\ttfamily]
method Max(a: int, b: int) returns (m: int)
  requires true
  ensures m >= a && m >= b
  ensures m == a || m == b
{
  if a >= b { m := a; } else { m := b; }
}
\end{lstlisting}
\caption{A simple Dafny method with specifications.}
\label{fig:dafny-example}
\end{figure}

The verification mechanism relies on \textit{weakest precondition} calculus. Dafny translates code and specifications into an intermediate representation (Boogie \citep{boogie}), which computes the weakest precondition $wp(s, Q)$—the necessary and sufficient condition for $s$ to terminate in a state satisfying $Q$. The system then generates a verification condition of the form $P \implies wp(s, Q)$, which is dispatched to the Z3 SMT solver \citep{z3}. If Z3 determines the condition is valid, the program is verified.

When Z3 cannot verify a program automatically, users provide \textit{ghost code}—annotations that guide the solver without affecting runtime behavior. Common forms include assertions (\texttt{assert P}) that establish intermediate facts, loop invariants that characterize state at each iteration, lemmas that encapsulate reusable proof steps, and calculational proofs (\texttt{calc}) that break complex equalities into transitive steps. Unlike interactive theorem provers where users construct explicit proof terms step-by-step, Dafny's SMT solver handles low-level reasoning automatically while users focus on high-level proof structure.

Dafny has enabled large-scale verification efforts across industry and academia, including AWS's Cedar authorization engine \citep{cedar}, the AWS Cryptographic Material Providers Library \citep{aws-crypto}, VMware's VeriBetrFS file system \citep{veribetrkv}, Microsoft's IronFleet distributed systems \citep{ironfleet}, and ConsenSys's Ethereum Virtual Machine specification \citep{dafny-evm}. Notably, AWS rewrote and verified its policy authorization engine in Dafny, which is invoked 1 billion times per second \citep{Chakarov2025}.

\section{ATLAS}

Synthesizing formally verified code poses challenges beyond standard code generation. A robust pipeline must produce both an implementation and the correctness specification it satisfies. Drawing from formal theorem proving principles, we identify three key requirements:

\begin{itemize}
\item \emph{Strong Specifications.} The pipeline must generate strong, semantically meaningful specifications that fully capture the program's intended logic—not merely safety properties like non-null outputs or range constraints.

\item \emph{Test Cases.} The pipeline must generate executable test cases within the formal language to ground specifications in concrete examples.

\item \emph{Problem Freezing.} Once defined, the problem statement must remain fixed to prevent the model from weakening specifications or trivializing test cases. This mirrors mathematical theorem proving, where a proof attempt cannot alter the theorem statement once finalized.

\end{itemize}

To address these requirements, we designed ATLAS, a multi-phase pipeline that decouples problem statement generation from verified solution synthesis. The complete process is shown in \Cref{fig:pipeline}.

\subsection{Overview}

\begin{figure}[t!]
\begin{lstlisting}[language=dafny, breaklines=true, frame=single]
predicate isStringPalindrome(s: string)
  ensures isStringPalindrome(s) <==> forall i :: 0 <= i < |s| / 2 ==> s[i] == s[|s| - 1 - i]
  ensures |s| <= 1 ==> isStringPalindrome(s)
{
  forall i :: 0 <= i < |s| / 2 ==> s[i] == s[|s| - 1 - i]
}

function reverseString(s: string): string
  ensures |reverseString(s)| == |s|
  ensures forall i :: 0 <= i < |s| ==> reverseString(s)[i] == s[|s| - 1 - i]
  ensures isStringPalindrome(s) <==> s == reverseString(s)
{
  if |s| == 0 then "" else reverseString(s[1..]) + [s[0]]
}
  
method IsPalindrome(s: string) returns (isPalindrome: bool)
  ensures isPalindrome == isStringPalindrome(s)
  ensures |s| <= 1 ==> isPalindrome
{
  isPalindrome := true;
  var i := 0;
  while i < |s| / 2
    invariant 0 <= i <= |s| / 2
    invariant isPalindrome == forall j :: 0 <= j < i ==> s[j] == s[|s| - 1 - j]
  {
    if s[i] != s[|s| - 1 - i] {
      isPalindrome := false;
      return;
    }
    i := i + 1;
  }
}

method Test() {
  // Test case 1 [TACO]
  var result1 := IsPalindrome("aba");
  expect result1 == true;
  ...
  // Test case 8 [GENERATED]
  var result8 := IsPalindrome("fnjzxnxnjplfwzowfdrk");
  expect result8 == false;
}
\end{lstlisting}
\caption{A verified Dafny program synthesized by ATLAS, including specification, implementation, and test cases.}
\label{fig:implementation}
\end{figure}

We seed the synthesis pipeline with a dataset of programming contest problems paired with solutions in a general-purpose language. This ensures each program has a self-contained implementation and reference solution with input-output test cases. We use TACO-verified \citep{li2023tacotopicsalgorithmiccode}, which contains 12.8K programming contest problems with Python reference solutions. We selected TACO-verified for its breadth—covering problems from LeetCode, CodeForces, HackerRank, and others—its unified and rich metadata, and its permissive licensing. Other datasets that could potentially be used for this task include APPS \citep{APPS} and CodeContests \citep{CodeConstests}, which offer similar problem-solution pairs. However, they may lack the same high-quality metadata and strict test-case quality that TACO-verified provides.

\subsubsection{Contract Generation}

The first stage generates a sound and complete formal contract—consisting of a method signature, strong pre- and postconditions, and executable test cases—that serves as the immutable problem statement for implementation.

\paragraph{Method Signatures.} We extract a reference solution from the dataset and prompt an LLM to generate the corresponding Dafny method signature. We use a generative approach to accommodate type system differences between Python and Dafny, a design choice validated by high task adherence rates.

\paragraph{Specifications and Tests.} Given the method signature, we synthesize specifications as \texttt{requires} and \texttt{ensures} clauses, which may reference auxiliary functions or predicates. In parallel, we generate test cases from the input-output examples in TACO-verified, supplemented with additional test cases generated by the agent to ensure broad coverage.

\paragraph{Iterative Refinement.} The Dafny compiler verifies generated specifications and tests for syntax errors. On failure, compiler errors are formatted and passed to an LLM judge, which provides corrective feedback for the next generation attempt. This iterative refinement loop continues until syntactically valid specifications and test cases are produced or the error-correction budget is exhausted. Beyond syntactic correctness, we apply soundness and completeness checks that significantly improve compatibility between test cases and specifications (detailed in \Cref{sec:soundness_and_completeness}).

\subsubsection{Implementation Generation}

Once a valid formal contract is generated and frozen, the second stage synthesizes implementations, assertions, and lemmas that provably satisfy the contract.

\paragraph{Code and Proofs.} We prompt an LLM with the specification and test cases to generate an implementation. We observe that in practice, frontier models struggle to output only implementations and assertions when specifications contain multiple functions or predicates. We therefore allow the model to regenerate specifications during implementation, provided they satisfy strict adherence constraints defined in the next validation step.

\paragraph{Validation.} We enforce three checks to ensure correctness. First, we verify contract adherence: each \texttt{predicate}, \texttt{function}, and \texttt{method} must have exactly the original \texttt{requires} clauses and a superset of the original \texttt{ensures} clauses. Second, we verify that the complete Dafny program satisfies the specification. Third, we compile the program and confirm that it passes all test cases in the \texttt{Test()} method.

An example of a program that successfully passed both phases is shown in \Cref{fig:implementation}.

\subsection{Soundness and Completeness}
\label{sec:soundness_and_completeness}

\begin{figure}[t!]
\begin{lstlisting}[frame=single,language=dafny, breaklines=true]
lemma Soundness(s: string, isPalindrome: bool)
  requires s == "aba" // Test input
  requires isPalindrome == true // Test output
  ensures isPalindrome == isStringPalindrome(s)
  ensures |s| <= 1 ==> isPalindrome
{}

lemma CompletenessContr(s: string, isPalindrome: bool)
  requires s == "aba" // Test input
  requires isPalindrome != true // Test output negated
  requires isPalindrome == isStringPalindrome(s)
  requires |s| <= 1 ==> isPalindrome
  ensures false
{}

lemma CompletenessPerturb(s: string, isPalindrome: bool)
  requires s == "aba" // Test input
  requires isPalindrome == false // Test output perturbed
  ensures isPalindrome == isStringPalindrome(s)
  ensures |s| <= 1 ==> isPalindrome
{}
\end{lstlisting}
\caption{Soundness and completeness lemmas constructed from the first test case in \Cref{fig:implementation}.}
\label{fig:soundness_and_completeness}
\end{figure}

During specification generation, we observe that synthesized specifications commonly exhibit insufficient strength, particularly in postconditions. This weakness is exacerbated by iterative error corrections, which progressively weaken postconditions rather than strengthen them. To address this, we develop a framework that systematically relates specifications and test cases through verification lemmas.

We construct three types of verification lemmas from test case input-output pairs:

\paragraph{Soundness Lemmas}
These lemmas evaluate whether the contract is consistent with test cases by checking if it holds pointwise for concrete instantiations. The intuition is that if a specification correctly captures the intended behavior, the contract should be satisfied when instantiated with known correct examples. For each test case, we construct a lemma that asserts the concrete input and output values, then verifies whether the contract holds (\Cref{fig:soundness_and_completeness}). If this lemma verifies successfully, it confirms that the contract is consistent with the test case. Conversely, if a soundness lemma fails, it reveals an inconsistency—the contract cannot be satisfied by this known correct example, indicating it is either too restrictive or incorrectly specified.

\paragraph{Completeness Lemmas (Contradiction)}
These lemmas evaluate whether the postconditions uniquely determine outputs for given inputs by attempting to derive a contradiction from alternative outputs. The intuition is that if a specification is sufficiently complete, assuming a different output while maintaining the contract should lead to an inconsistency. For each test case, we construct a lemma that assumes the preconditions, postconditions, and negation of the expected output, then attempts to derive false (\Cref{fig:soundness_and_completeness}). If this lemma proves successfully, it confirms no other output can satisfy the postconditions for the given input. However, such proofs often exceed automated theorem provers capabilities without guidance, motivating our perturbation-based alternative described in the next paragraph.

\paragraph{Completeness Lemmas (Perturbation)}
These lemmas provide a practical alternative for evaluating specification completeness by directly testing whether the postconditions accept incorrect outputs. While the contradiction approach attempts to prove that alternative outputs are impossible—a task often too complex for automated provers—the perturbation approach simply checks whether modified outputs satisfy the contract. For each test case, we systematically perturb the output by structurally prompting an LLM while maintaining fixed inputs, then verify whether the contract still holds (\Cref{fig:soundness_and_completeness}). If a lemma verifies successfully with a perturbed output, it reveals that the postconditions have insufficient constraints—they accept behavior that should be rejected. Conversely, if the perturbed output fails verification, it suggests the postconditions are appropriately restrictive. This approach trades formal completeness guarantees for practical automated verification. 

We leverage Dafny's auto-active verification to obtain prover feedback on these lemmas without explicit proof construction. The verification results inform the LLM judge following specification generation, enabling accurate evaluation of specification strength and consistency. Failed soundness lemmas indicate inconsistent or overly restrictive contracts, while successful perturbation lemmas reveal incomplete specifications. While this approach is limited by the base prover's capabilities, it significantly improves specification quality by detecting weaknesses that would otherwise pass syntactic validation alone.

\subsection{Analysis}

Applied to TACO-verified, ATLAS successfully generates 2,751 verified Dafny programs, each with specifications and machine-checked proofs. We analyze pipeline performance across difficulty levels and problem types.

\begin{figure}[t]
    \centering
      \resizebox{\columnwidth}{!}{%
  \begin{tabular}{lrrrrr}
    \toprule
    \textbf{Difficulty} & \textbf{Total} & \textbf{Success} & \textbf{Success} & \textbf{Avg. Success} & \textbf{Avg. Fail} \\
    & \textbf{Samples} & \textbf{Samples} & \textbf{Rate (\%)} & \textbf{Length} & \textbf{Length} \\
    \midrule
    EASY & 5456 & 2572 & 47.14 & 3848 & 6531 \\
    MEDIUM & 1353 & 338 & 24.98 & 4902 & 6879 \\
    MEDIUM HARD & 1801 & 451 & 25.04 & 5226 & 6929 \\
    HARD & 1256 & 260 & 20.70 & 5692 & 7270 \\
    VERY HARD & 537 & 107 & 19.93 & 5637 & 7102 \\
    \midrule
    UNKNOWN & 2000 & 689 & 34.45 & 4559 & 6929 \\
    \bottomrule
  \end{tabular}%
  }\\
  \caption{ATLAS pipeline success rate by TACO-verified difficulty ratings.}
    \label{fig:success_rate}
\end{figure}

\begin{figure}[t]
    \centering
\includegraphics[width=\columnwidth]{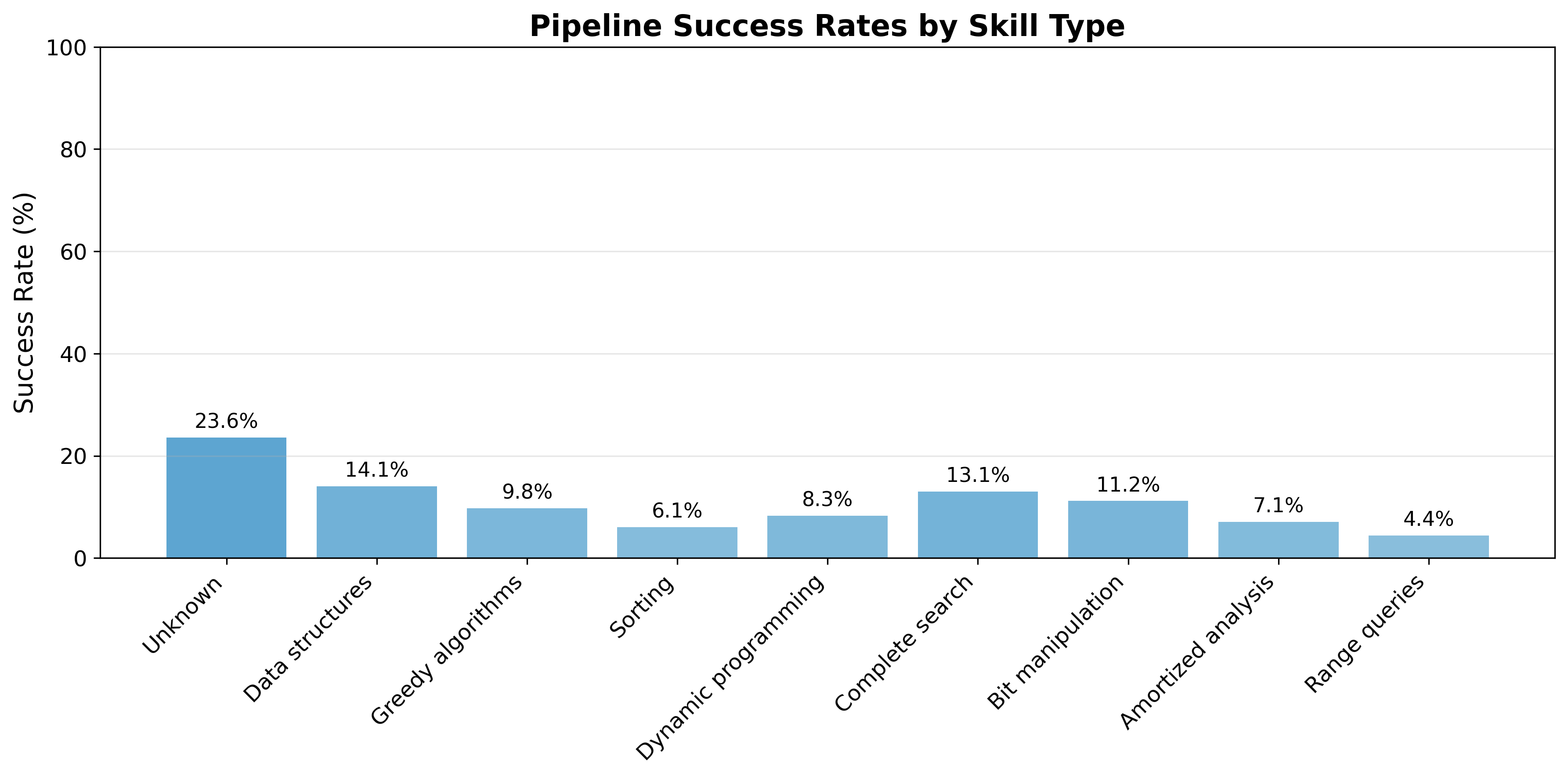}
  \caption{ATLAS pipeline success rate by TACO-verified difficulty skill types.}
    \label{fig:success_skill_sets}
\end{figure}

\subsubsection{Difficulty Ratings}
To evaluate the robustness of our ATLAS pipeline, we analyzed its performance against the difficulty ratings provided by the TACO-verified dataset. The dataset classifies problems into five tiers—\texttt{EASY}, \texttt{MEDIUM}, \texttt{MEDIUM HARD}, \texttt{HARD}, and \texttt{VERY HARD}—enabling us to measure success rate and output characteristics as a function of problem complexity. Results are shown in \Cref{fig:success_rate}

The data shows a clear inverse correlation between problem difficulty and success rate. Success rate is highest for \texttt{EASY} problems at 47.1\% and decreases systematically with complexity, reaching approximately 20\% for \texttt{HARD} and \texttt{VERY HARD} problems. This validates that difficulty ratings meaningfully reflect synthesis and verification challenge.

The average length of successful programs increases consistently with difficulty, growing from 3,848 characters for \texttt{EASY} problems to over 5,600 for the hardest categories, suggesting that complex problems require more elaborate code and proof structures. Conversely, failed attempts remain consistently long across all categories, indicating that failures stem from complex, verbose, but ultimately incorrect generations rather than trivial attempts. These patterns confirm the rational behavior of our synthesis pipeline.

\subsubsection{Skill Set}

TACO-verified provides skill set tags for many problems, enabling granular analysis of synthesis capabilities presented in \Cref{fig:success_skill_sets}. Among tagged problems, data structure problems achieve the highest success rate at 14.1\%, followed by search problems at 13.1\%. The pipeline struggles more with sorting and range query problems (6.1\% and 4.4\%, respectively). Range query problems involve efficiently answering queries about intervals of objects (e.g., determining the maximum sum of a contiguous subarray of fixed length $k$ within a given array), whereas higher-success problem types typically require manipulation of individual elements (e.g., finding the closest-valued node with exactly $k$ descendants). This pattern suggests that formally verifying operations on groups of objects presents a greater challenge than verifying discrete element manipulations, though comprehensive studies would be necessary to substantiate this observation.

\section{Supervised Fine-Tuning}

The verified programs produced by ATLAS provide a foundation for training models on formal verification. Rather than using only final outputs, we extract multiple training tasks from each pipeline run—including intermediate attempts and repairs—to maximize data efficiency. We describe our dataset curation, training procedure, and evaluation on two Dafny benchmarks.

\subsection{Training}

\subsubsection{Dataset Curation}

\begin{table}[t]
  \caption{Multi-task SFT dataset produced by ATLAS.}
  \label{tab:sft_all_tasks}
  \centering
  \renewcommand{\arraystretch}{1.2}

  \begin{tabularx}{\columnwidth}{@{} p{5.5cm} >{\raggedright\arraybackslash}X r r @{}}
    \toprule
    \textbf{Task} & \textbf{Count} & \textbf{(\%)} \\
    \midrule

    NL-to-Code Synthesis
    & 2,751 & 14.19 \\
    \addlinespace

    NL-to-Spec Synthesis
    & 5,225 & 14.19 \\
    \addlinespace

    Spec-to-Code Synthesis 
    & 2,751 & 14.19 \\
    \addlinespace

    Specification Repair
    & 2,353 & 12.14 \\
    \addlinespace

    Implementation Repair 
    & 4,562 & 23.53 \\
    \addlinespace

    Proof Infilling
    & 1,743 & 8.99 \\
    \addlinespace

    \midrule
    \textbf{Total} & \textbf{19,385} & \textbf{100.00} \\
    \bottomrule
  \end{tabularx}
\end{table}

We hypothesize that fine-tuning for formal verification requires a dataset that teaches more than final answers \citep{cot-reasoning}. For agentic applications, the training data must capture the process of problem decomposition, specification, implementation, and correction. We therefore train not only on the final verified Dafny programs ATLAS produces, but also on its execution traces, including intermediate failed attempts. This approach significantly increases the per-sample efficiency of each verified program.

The curated dataset includes: the initial natural language prompt, all intermediate generation attempts for specifications and implementations, LLM-as-a-Judge feedback, and the final verified program. As shown in \Cref{tab:sft_all_tasks}, this enables us to extract multiple training tasks from each pipeline run: direct implementation (NL-to-Code Synthesis), specification formalization (NL-to-Spec Synthesis), implementation from specifications (Spec-to-Code Synthesis), assertion synthesis (Proof Infilling), and correction tasks. Correction tasks comprise over 35\% of our dataset (Specification Repair and Implementation Repair combined), transforming typically discarded failed attempts into valuable learning opportunities. The Proof Infilling task further leverages successful verifications by teaching the model to reconstruct essential proof annotations such as loop invariants and auxiliary lemmas. These examples are created by stripping proof hints from verified programs. Of the 2.7K verified programs produced by ATLAS, approximately 1K verify automatically without hints, yielding 1.7K training examples.

This multi-task decomposition yields over 7 training examples per verified program on average, significantly amplifying the utility of each successful pipeline run.

\subsubsection{Execution}
This multi-task approach structures our pipeline output into a curriculum for learning formal verification. We fine-tune Qwen 2.5 7B Coder, an open-source LLM for general coding that has limited knowledge of verification-aware languages and syntax. We perform LoRA \citep{hu2021loralowrankadaptationlarge} supervised fine-tuning using the Hugging Face TRL library \citep{trl} on an AWS EC2 instance with 8 NVIDIA H100 GPUs. The model trains for up to 10 epochs using standard cross-entropy loss with gradient clipping. We evaluate on two downstream Dafny benchmarks to assess whether the ATLAS dataset enhances verified code synthesis capabilities.

\vspace{1cm}
\subsection{Evaluation} 

\subsubsection{DafnyBench}

\begin{figure}[t!]
    \centering
    \includegraphics[width=\columnwidth]{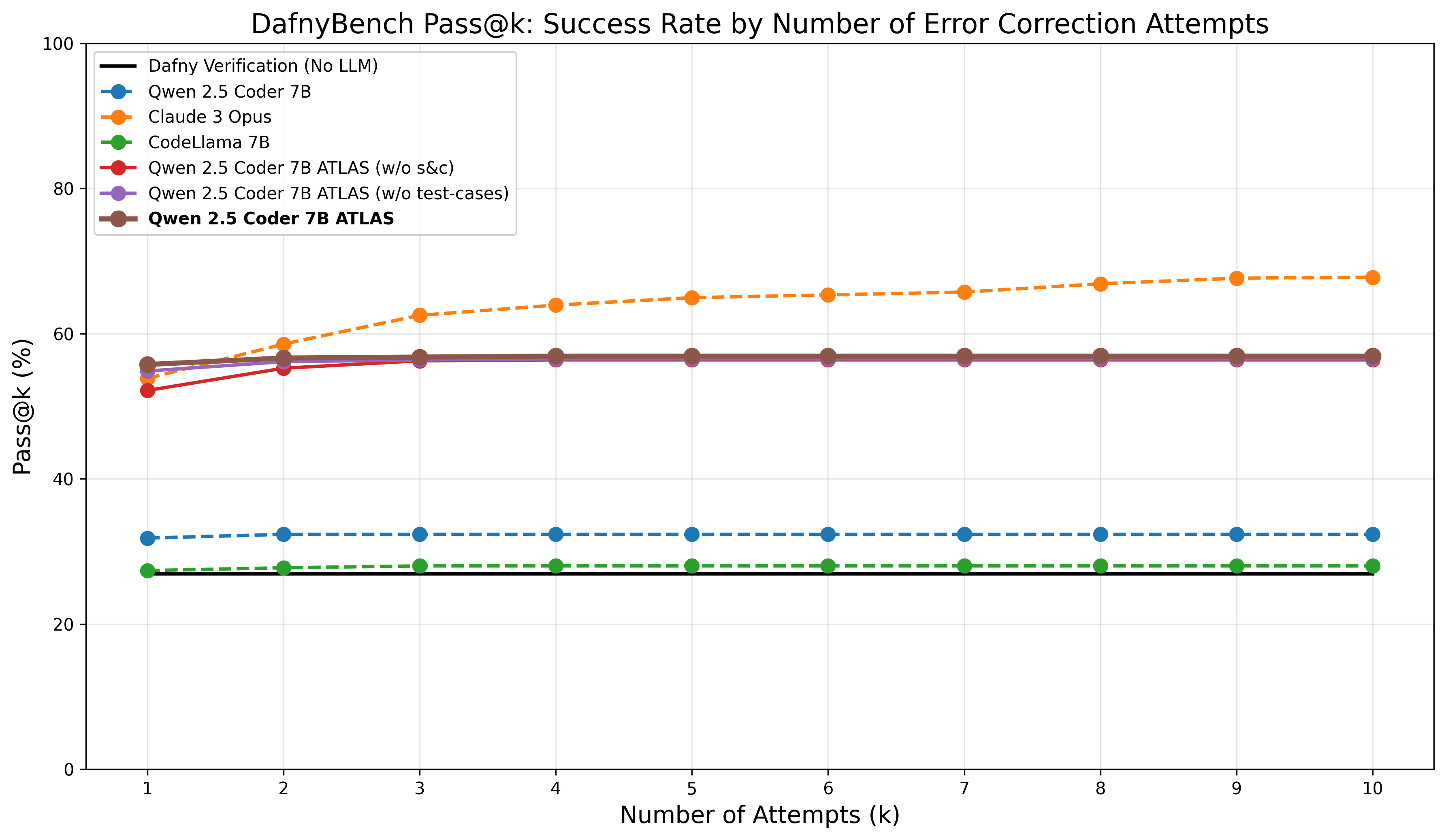}\\
\small
\begin{tabular}{@{}lrrr@{}}
\toprule
\textbf{Model} & \textbf{Pass@1} & \textbf{Pass@5} & \textbf{Pass@10} \\
\midrule
Claude 4.1 Opus\footnotemark[1] & -- & -- & 89.2 \\
GPT-5\footnotemark[1] & -- & -- & 72.0 \\
Gemini 2.5 Pro\footnotemark[1] & -- & -- & 75.3 \\
\midrule
Claude 3 Opus\footnotemark[2] & 53.8 & 65.0 & 67.8 \\
CodeLlama 7B\footnotemark[2] & 27.4 & 28.0 & 28.0 \\
\midrule

Dafny Verification (No LLM) & 26.9 & 26.9 & 26.9 \\
Qwen 2.5 Coder 7B & 31.8 & 32.4 & 32.4 \\
\textbf{ATLAS Qwen 2.5 Coder 7B} & \textbf{55.8} & \textbf{56.9} & \textbf{56.9} \\
\quad w/o s\&c & 52.2 & 56.4 & 56.4 \\
\quad w/o test cases & 54.9 & 56.4 & 56.5 \\

\bottomrule
\end{tabular}
    \caption{Evaluation results on DafnyBench.}
\label{tab:dafnybench_results}

\end{figure}

\footnotetext[1]{As reported in \citep{vericoding}.}
\footnotetext[2]{As reported in \citep{loughridge2024dafnybenchbenchmarkformalsoftware} at the time of benchmark release.}

To assess generalization, we evaluate our fine-tuned model on DafnyBench \citep{loughridge2024dafnybenchbenchmarkformalsoftware}, a widely recognized benchmark for Dafny proof hint generation. The benchmark comprises 782 problems requiring assertions and invariants to be filled into ground-truth implementations with specifications.

As shown in \Cref{tab:dafnybench_results}, ATLAS-enhanced Qwen 2.5 Coder 7B achieves strong single-shot performance at 55.8\% Pass@1, surpassing Claude 3 Opus (53.8\%), the state-of-the-art at the benchmark's release \citep{loughridge2024dafnybenchbenchmarkformalsoftware}. However, performance plateaus at Pass@5 (56.9\%), showing minimal improvement with additional attempts. This contrasts with Claude models, which benefit significantly from multiple attempts—Claude 3 Opus improves from 53.8\% to 67.8\%. Recent frontier models have established higher performance \citep{vericoding}, with Opus 4.1 achieving 89.2\% at Pass@10 \citep{vericoding}, suggesting our approach may be constrained by the 7B parameter scale of our base model.

Ablation studies reveal that removing specification and completeness checks or test cases has minimal impact on DafnyBench performance, with Pass@10 scores remaining around 56\%. This is expected, as DafnyBench primarily focuses on implementing pre-defined contracts rather than open-ended synthesis tasks where these verification components typically prove more valuable (see \Cref{section:DafnySynthesis}). The benchmark's nature—having ground-truth specifications and not requiring test case generation—likely undermines the performance boost of soundness and completeness (s\&c) checks and test case generation compared to open-ended synthesis tasks. The substantial improvement over the base model Qwen 2.5 Coder 7B (32.4\% at Pass@10) and size-matched CodeLlama 7B (28.0\% at Pass@10) demonstrates that our ATLAS dataset successfully induces verification-aware capabilities, even if scaling limitations prevent matching larger frontier models.

\subsubsection{DafnySynthesis}
\label{section:DafnySynthesis}

\begin{figure}[t]
    \centering
    \includegraphics[width=\columnwidth]{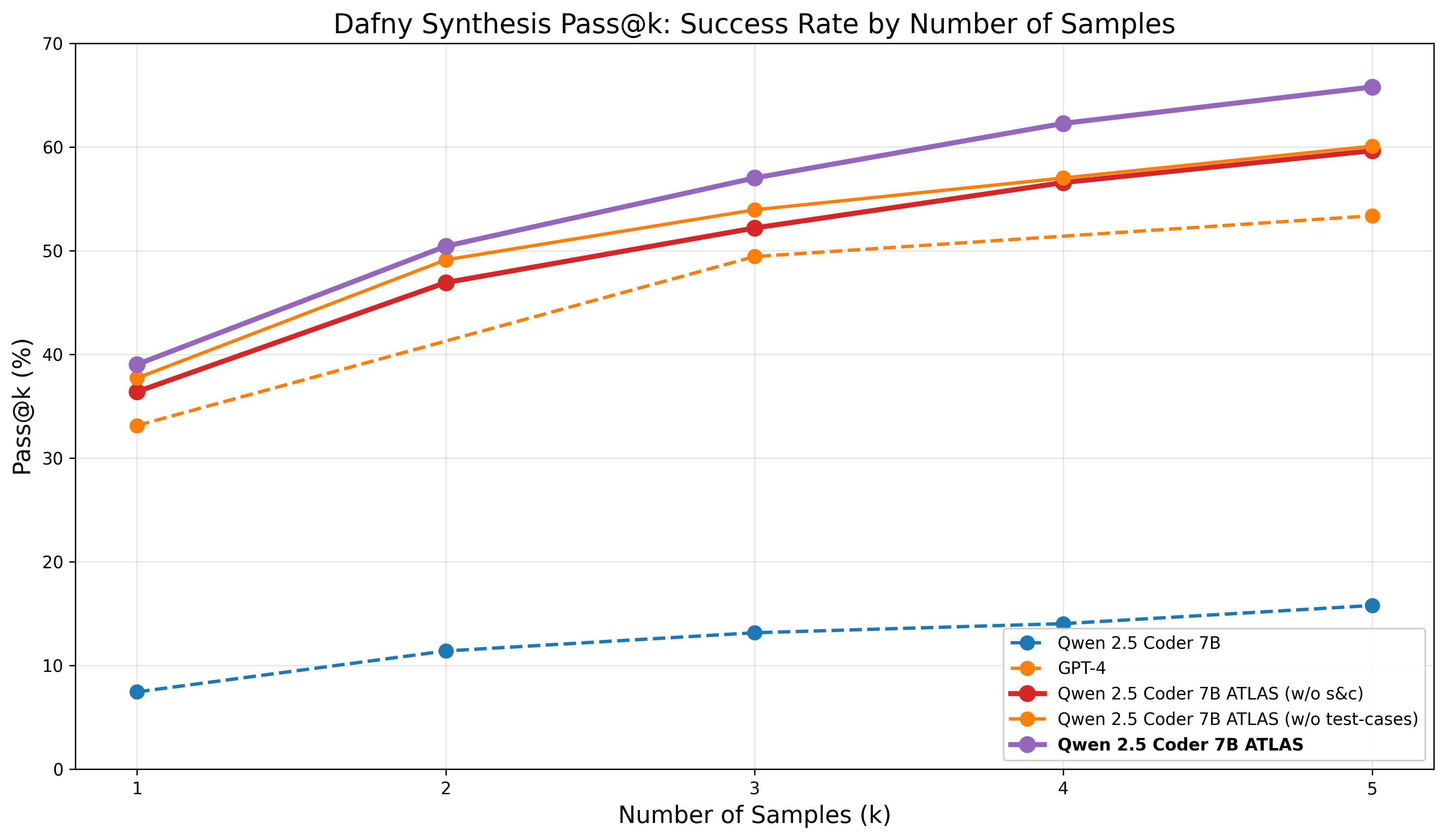}\\
\small
\begin{tabular}{@{}lrrr@{}}
\toprule
\textbf{Model} & \textbf{Pass@1} & \textbf{Pass@3} & \textbf{Pass@5} \\
\midrule
Qwen 2.5 Coder 7B & 7.5 & 13.2 & 15.8 \\
GPT-4\footnotemark[1] & 33.1 & 49.4 & 53.4 \\
\textbf{ATLAS Qwen 2.5 Coder 7B} & \textbf{39.0} & \textbf{57.0} & \textbf{65.8} \\
\quad w/o s\&c & 36.4 & 52.2 & 58.4 \\
\quad w/o test cases & 32.6 & 51.7 & 59.6\\
\bottomrule
\end{tabular}
\caption{Evaluation results on DafnySynthesis.}
\label{tab:dafnysynthesis_results}
\end{figure}

To evaluate our model on Dafny code synthesis, we use the DafnySynthesis benchmark \citep{DafnySynthesis}, which consists of 228 programming tasks adapted for Dafny verification split into a test set of 178 tasks and 50 tasks for training. This dataset derives from the sanitized MBPP (Mostly Basic Python Programming) dataset \citep{mbpp}, curated to exclude problems requiring Python-specific features or external libraries incompatible with Dafny's standard libraries. Each problem includes a natural language description of the task, Dafny method signature, and 3 test cases. The evaluation of the benchmark considers a response successful if it verifies without errors and contains non-trivial specifications of the functions or methods synthesized.

We evaluate our model using the signature prompt format from \citep{DafnySynthesis}, which combines the problem description with its method signature and test cases. While \citep{DafnySynthesis} explored contextless and Chain of Thought (CoT) prompting, we selected the signature format as it provides a standardized and practical assessment of verification-aware programming, balancing realistic context with minimal scaffolding.

Results are shown in \Cref{tab:dafnysynthesis_results}. ATLAS-trained Qwen 2.5 Coder 7B achieves 65.8\% verification success at Pass@5, surpassing GPT-4 (53.4\%) by 12.4 percentage points, which was the best performing model at benchmark release. The substantial improvement over the baseline model Qwen 2.5 Coder 7B (15.8\% at Pass@5) demonstrates the effectiveness of the ATLAS dataset in inducing verification-aware programming capabilities.

Ablation studies reveal the contributions of different components to ATLAS performance. Removing soundness and completeness checks (s\&c) degrades Pass@5 to 58.4\%, indicating these components are crucial for guiding the model toward formally verified solutions. Similarly, removing test cases from fine-tuning yields 59.6\% at Pass@5, demonstrating the value of concrete examples in the synthesis process.

\footnotetext[1]{As reported in \citep{DafnySynthesis} at the time of benchmark release.}
\section{Related and Future Work} 

Complementary to our work, dafny-annotator \citep{dafny-annotator} targets automated annotation generation by bootstrapping from existing Dafny programs. In contrast, ATLAS employs a stage-wise approach that generates specifications before implementations without requiring Dafny code as input. While dafny-annotator evaluates on a test split of DafnyBench after using the remainder for synthesis and training, ATLAS achieves strong performance on DafnyBench without in-distribution training, demonstrating that general-purpose synthesis can effectively support specialized verification tasks.

Several promising research directions remain unexplored. First, while current benchmarks evaluate free-form synthesis or specialized tasks like proof hint infilling, there is a need for benchmarks emphasizing end-to-end synthesis for complete specifications. Recent work such as \citep{vericoding}, VERINA \citep{verina}, and CLEVER \citep{clever} in Lean represent progress in this direction.

Second, while reinforcement learning has shown success in mathematical theorem proving \citep{wang2025kiminaproverpreviewlargeformal, lin2025goedelproverfrontiermodelopensource}, its application to verification-aware code generation remains underexplored due to lack of well-defined reward structures. Future work should investigate combining compiler feedback, complexity analysis, and proof resource allocation as reward signals.

Finally, recent advances in agentic theorem proving, including HILBERT \citep{hilbert} and SeedProver \citep{seed-prover}, suggest promising directions. These approaches demonstrate how agentic interaction with verification frameworks can augment model capabilities through iterative refinement and guided exploration, with potential for practical co-pilot and agent tools.

\section{Conclusion}

Our work ATLAS, an automated pipeline for synthesizing verified Dafny programs from algorithmic problems with reference implementations. By introducing soundness and completeness checks that leverage test cases, ATLAS ensures specification quality and filters degenerate solutions. Applied to TACO-verified, the pipeline produces 2.7K verified programs, which we decompose into 19K training examples spanning synthesis, repair, and proof infilling tasks. Fine-tuning Qwen 2.5 7B Coder on this data yields substantial improvements on both DafnyBench (32.4\% to 56.9\%) and DafnySynthesis (15.8\% to 65.8\%), demonstrating that synthetic data generation is a viable path to scaling LLM capabilities for formal verification. Our results suggest that the data scarcity bottleneck for verification-aware languages can be addressed through automated synthesis pipelines, opening opportunities for broader adoption of formal methods in software development.

\section*{Impact Statement}
This paper presents work whose goal is to advance the field of Machine Learning. There are many potential societal consequences of our work, none which we feel must be specifically highlighted here.
\bibliography{bibliography}
\bibliographystyle{icml2026}

\newpage
\appendix
\onecolumn
\section{Appendix}

\section*{Code Examples}

\subsection*{Incomplete Specification and Implementation}

The listing below demonstrates a naive implementation of an arithmetic progression checker produced by ATLAS without soundness and completeness checks. While syntactically correct, it lacks crucial specification properties. The implementation only checks consecutive differences, missing ensures clauses specifying the relationship between all pairs of adjacent elements.

\begin{lstlisting}[frame=single,language=dafny, breaklines=true]
method ArithmeticProgression(sequence: seq<int>) 
  returns (hasSamePattern: bool)
    requires |sequence| >= 3
{
    var diff := sequence[1] - sequence[0];
    hasSamePattern := (forall i :: 0 <= i < |sequence|-1 ==> 
        sequence[i+1] - sequence[i] == diff);
    return;
}
\end{lstlisting}

\subsection*{Complete Specification and Implementation using Soundness and Completeness checks}

The addition of soundness and completeness checks results in output, which demonstrates robust and complete specifications. In the below listing, the solution introduces two auxiliary functions: \texttt{IsArithmeticSequence} establishes the core property that all adjacent differences must be equal, while \texttt{IsArithmeticProgression} adds the size requirement and connects to the main implementation. 

\begin{lstlisting}[frame=single,language=dafny, breaklines=true]
function IsArithmeticSequence(sequence: seq<int>): bool
  requires |sequence| >= 2
  ensures IsArithmeticSequence(sequence) <==>
    forall i, j :: 0 <= i < j < |sequence|-1 ==> 
      (sequence[i+1] - sequence[i] == 
       sequence[j+1] - sequence[j])
{
  if |sequence| < 2 then false
  else
    var diff := sequence[1] - sequence[0];
    forall i :: 0 <= i < |sequence|-1 ==> 
        sequence[i+1] - sequence[i] == diff
}

function IsArithmeticProgression(
    sequence: seq<int>): bool
  requires |sequence| >= 3
  ensures IsArithmeticProgression(sequence) <==> 
    (|sequence| >= 3) && 
    IsArithmeticSequence(sequence)
{
  |sequence| >= 3 && IsArithmeticSequence(sequence)
}

method ArithmeticProgression(sequence: seq<int>) 
    returns (hasSamePattern: bool)
  requires |sequence| >= 3
  ensures hasSamePattern <==> 
    IsArithmeticProgression(sequence)
{
  var diff := sequence[1] - sequence[0];
  var i := 1;
  hasSamePattern := true;
  
  while i < |sequence| - 1
    invariant 1 <= i <= |sequence| - 1
    invariant hasSamePattern <==> 
      (forall k :: 0 <= k < i ==> 
        sequence[k+1] - sequence[k] == diff)
  {
    if sequence[i+1] - sequence[i] != diff {
      hasSamePattern := false;
      return;
    }
    i := i + 1;
  }
}
\end{lstlisting}

\section*{Soundness and Completeness Lemma Generation}
\label{app:lemma_generation}

Algorithm~\ref{alg:lemma_gen} presents the procedure for constructing verification lemmas that evaluate specification quality. The algorithm takes as input a method $m$ with its signature $\sigma$, pre-conditions $R$, post-conditions $E$, and a test suite $\mathcal{T}$ containing $n$ test cases. Each test case $(a_i, o_i, \Gamma_i)$ comprises input arguments $a_i$, expected output $o_i$, and input variable bindings $\Gamma_i$ that map method parameters to concrete values.

\paragraph{Soundness Lemmas.} We instantiate the contract with concrete test inputs and outputs as preconditions, then verify that the postconditions hold. Verification failure indicates the specification is inconsistent with known correct behavior.

\paragraph{Contradiction Lemmas.} We assume the contract holds with a negated output and attempt to derive false. Success proves the postconditions uniquely determine the output, though such proofs often exceed automated prover capabilities.

\paragraph{Perturbation Lemmas.} We substitute LLM-generated incorrect outputs—such as negated booleans, offset integers, or sequences with swapped elements—and check if the contract still holds. Successful verification reveals the specification admits incorrect behavior and requires strengthening.

\paragraph{Feedback Integration.} The complete lemma set $\mathcal{L}$ is submitted to the Dafny verifier. Verification results are parsed and formatted into structured feedback:

\begin{lstlisting}[frame=single, language=dafny, breaklines=true, basicstyle=\small\ttfamily]
Soundness Check Results:
  Test 1: PASS (contract consistent with test case)
  Test 2: FAIL (contract rejects valid input-output pair)
  
Completeness Check Results:
  Test 1: PASS (perturbed output correctly rejected)
  Test 2: FAIL (perturbed output incorrectly accepted)
\end{lstlisting}

This feedback is provided to the LLM Judge, which synthesizes actionable recommendations for specification refinement. Common remediation patterns include strengthening postconditions with additional constraints, introducing auxiliary predicates to capture invariant properties, and adding quantified assertions over collection elements.

\begin{algorithm}[ht]
   \caption{Soundness and Completeness Lemma Generation}
   \label{alg:lemma_gen}
\begin{algorithmic}[1]
   \STATE {\bfseries Input:} Method $m$, signature $\sigma$, pre-conditions $R$, post-conditions $E$, test cases $\mathcal{T} = \{(a_i, o_i, \Gamma_i)\}_{i=1}^n$ ($a_i$: inputs, $o_i$: outputs, $\Gamma_i$: input variable bindings)
   \STATE {\bfseries Output:} Set of verification lemmas $\mathcal{L}$
   \STATE $\mathcal{L} \gets \emptyset$
   \FOR{$i = 1$ {\bfseries to} $n$}
      \STATE $(a_i, o_i, \Gamma_i) \gets \mathcal{T}_i$
      \STATE // Soundness lemma
      \STATE $R_i \gets R \cup \Gamma_i \cup \{\sigma = a_i\} \cup \{v_{\text{out}} = o_i\}$
      \STATE $\ell^{\text{sound}}_i \gets \text{ConstructLemma}(\sigma, v_{\text{out}}, R_i, E)$
      \STATE $\mathcal{L} \gets \mathcal{L} \cup \{\ell^{\text{sound}}_i\}$
      \STATE // Completeness lemma (contradiction)
      \STATE $R_i \gets R \cup \Gamma_i \cup \{\sigma = a_i\} \cup \{v_{\text{out}} \neq o_i\} \cup E$
      \STATE $\ell^{\text{contra}}_i \gets \text{ConstructLemma}(\sigma, v_{\text{out}}, R_i, \{\bot\})$
      \STATE $\mathcal{L} \gets \mathcal{L} \cup \{\ell^{\text{contra}}_i\}$
      \STATE // Completeness lemma (perturbation)
      \STATE $p_i \gets \text{Perturbation}(o_i, E)$
      \STATE $R_i \gets R \cup \Gamma_i \cup \{\sigma = a_i\} \cup \{v_{\text{out}} = p_i\}$
      \STATE $\ell^{\text{perturb}}_i \gets \text{ConstructLemma}(\sigma, v_{\text{out}}, R_i, E)$
      \STATE $\mathcal{L} \gets \mathcal{L} \cup \{\ell^{\text{perturb}}_i\}$
   \ENDFOR
   \STATE {\bfseries return} $\mathcal{L}$
\end{algorithmic}
\end{algorithm}

\end{document}